\documentclass[11pt]{article}
\usepackage{moriond,epsfig}

\def\be{\begin{equation}}
\def\ee{\end{equation}}
\def\bea{\begin{eqnarray}}
\def\eea{\end{eqnarray}}

\newcommand{\numu}{$\nu_{\mu}$} 
\newcommand{\nue}{$\nu_{e}$}
 
\newcommand{\nutau}{$\nu_{\tau}$}
\newcommand{\dmsq}{$\Delta \rm{m}^{2}$}
\newcommand{\sintwo}{sin$^{2}$2$\theta$}

\newcommand{\numutonutau}{$\nu_{\mu}\rightarrow\nu_{\tau}$}
\newcommand{\numutonue}{$\nu_{\mu}\rightarrow\nu_{e}$}
\newcommand{\numutonust}{$\nu_{\mu}\rightarrow\nu_{s}$}

\begin{document}
\vspace*{4cm}
\title{LONG BASELINE NEUTRINO OSCILLATION REVIEW}

\author{ A.\ WEBER }

\address{Department of Physics, University of Oxford, Keble Road,\\
Oxford OX1 3RH, England}

\maketitle\abstracts{
This article will summaries the status of current and future long baseline
neutrino oscillation Experiments.
}

\section{Introduction}

The most exciting result of the last decade has been the growing
evidence that neutrinos are massive\cite{antonella}.  The V-A structure
of the electroweak theory requires only left handed and therefore
massless neutrinos.  Observation of neutrino mass is therefore in direct
conflict with the Standard Model.

The most important result, indicating that neutrinos do oscillate and
therefore have to have mass, comes from the Super Kamiokande
experiment. They observed a zenith angle dependence (and thus a
dependence on distance travelled) in the ratio of \numu\ to \nue\
interactions of atmospheric neutrinos \cite{superk-98}.  Since the
angular dependence is far larger than the asymmetry in the cosmic ray
flux, there is no known way of generating this effect other than neutrino
physics.  In addition the angular dependence is described
by oscillation parameters consistent with the measured
suppression of the total \numu\ to \nue\ ratio.  The angular
dependence at Super Kamiokande implies that the value of \dmsq, the
difference in the mass squared of the two neutrinos, lies between
$10^{-2}$  and $10^{-3}$eV$^{2}$.  The data from other experiments
confirms this observation\cite{atmos}.

\section{The Experiments}

There are now several experiments using man-made neutrino sources which
will try to confirm this observation and measure the 
parameters of neutrino oscillation. They use of man-made neutrino
source overcomes one of the main limitations of current experiments, 
the not well understood properties of the neutrino source.

\subsection{The K2K Experiment}

\begin{figure}
\centerline{\includegraphics[width=0.5\textwidth]{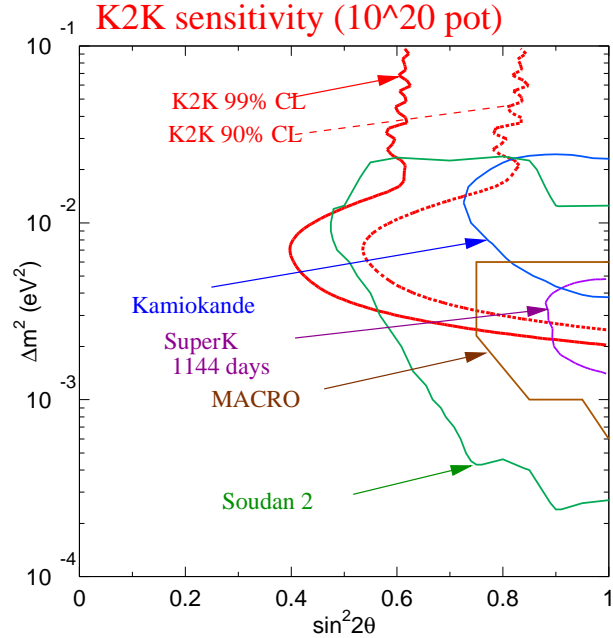}}
\caption{\label{fig:k2k} 
K2K expected sensitivity for $10^{20}$ protons on target, compared with
the allowed regions other experiments in oscillation parameters space.}
\end{figure}

The K2K experiment\cite{k2k}, which started taking data in 1999, studies
the region of \dmsq\ suggested by the atmospheric neutrino experiments.
It is the first long baseline neutrino oscillation experiment ever
built. An almost pure beam of $\bar{\nu}_\mu$ with an average energy
around 1.3 GeV is directed from KEK, Japan, to the SuperKamiokande
detector 250 km away. The beam energy spectrum and flux are measured
with a detector close to the neutrino production site and are than
extrapolated to estimate the expected flux and spectrum at the Kamioka
mine.  This experiment is sensitive only to some fraction of the region
of neutrino mass difference suggested by the atmospheric neutrino
results.  The beam intensity is not sufficient to cover the whole of the
Super-K allowed region (see Fig.\ \ref{fig:k2k}) and the beam energy is
too low to produce $\tau$ leptons in the preferred \numutonutau\
oscillation mode.

However, results presented at this conference\cite{k2k_moriond} already
show an impressive preliminary result using just the expected event
rate.  K2K has collected less than half of its expected data so far and
results will become more precise once the analysis of the neutrino
energy spectrum is included.

\subsection{The MINOS Experiment}

The MINOS experiment will study muon neutrinos produced at FNAL in a
controlled beam experiment.  A tunable beam of 1-20 GeV \numu\ will be sampled
before and after a journey of 735 km to the Soudan mine in Minnesota.
This baseline and choice of energy renders the experiment sensitive to
the entire region of neutrino mass difference suggested by the
atmospheric neutrino experiments.  MINOS will unambiguously confirm or
deny neutrino oscillation as the explanation of the atmospheric neutrino
anomaly.  If neutrinos do oscillate it will provide precise
measurements of the \dmsq\ and the mixing matrix elements.

There will be two functionally similar detectors: calorimeters made of
planes of 2.54 cm thick steel interleaved with sensitive planes
of 1cm thick x 4cm wide solid scintillator strips. The steel plates are
octagonal in shape and toroidally magnetised by a coil threaded
through a hole in the centre of the octagon.
In the far detector the regular octagons are 8 m in diameter and are grouped
into two ``supermodules''.  The total mass of the far detector is 5.4 ktons.
The far detector scintillator strips are read out at both ends by
wavelength shifting fibres onto Hamamatsu M16 multi-anode phototubes.
Eight strips are summed onto each pixel of the phototube.  The pattern
of summing at each end is arranged such that the strip which is crossed by a
single particle is uniquely determined. A more
detailed description can be found in \cite{m-tdr}.

The 0.98 kton near detector is offset with respect to the beam in order
to prevent 
the neutrino events from intersecting the coil hole. The squashed
octagonal shape of the detector allows the toroidal field to focus muons
towards the centre of the detector. Only the innermost 25cm of the beam will
be used in the analysis (the near and far beam spectra are most similar
for this region) and so only one quadrant needs to be instrumented in
the upstream region. The magnetic field has been designed such that the
average field seen by a muon in the near detector is as close as
possible to that seen by muons in the far detector.  The very large
number of neutrino interactions in the near detector opens the
possibility of high statistic conventional neutrino measurements.

\begin{figure}
\centerline{
\includegraphics[width=0.59\linewidth]{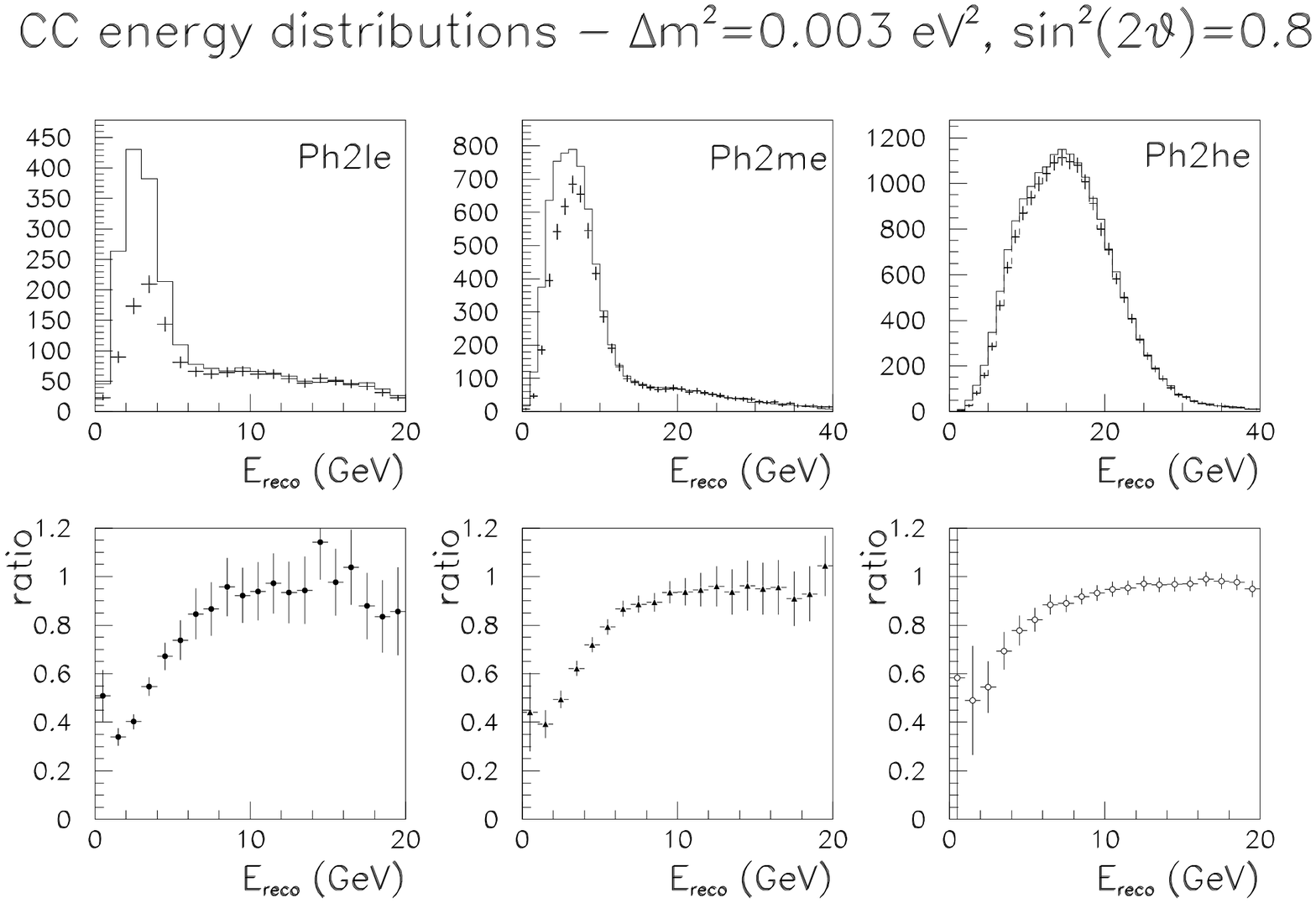}
\hspace{0.1cm}
\includegraphics[width=0.39\linewidth]{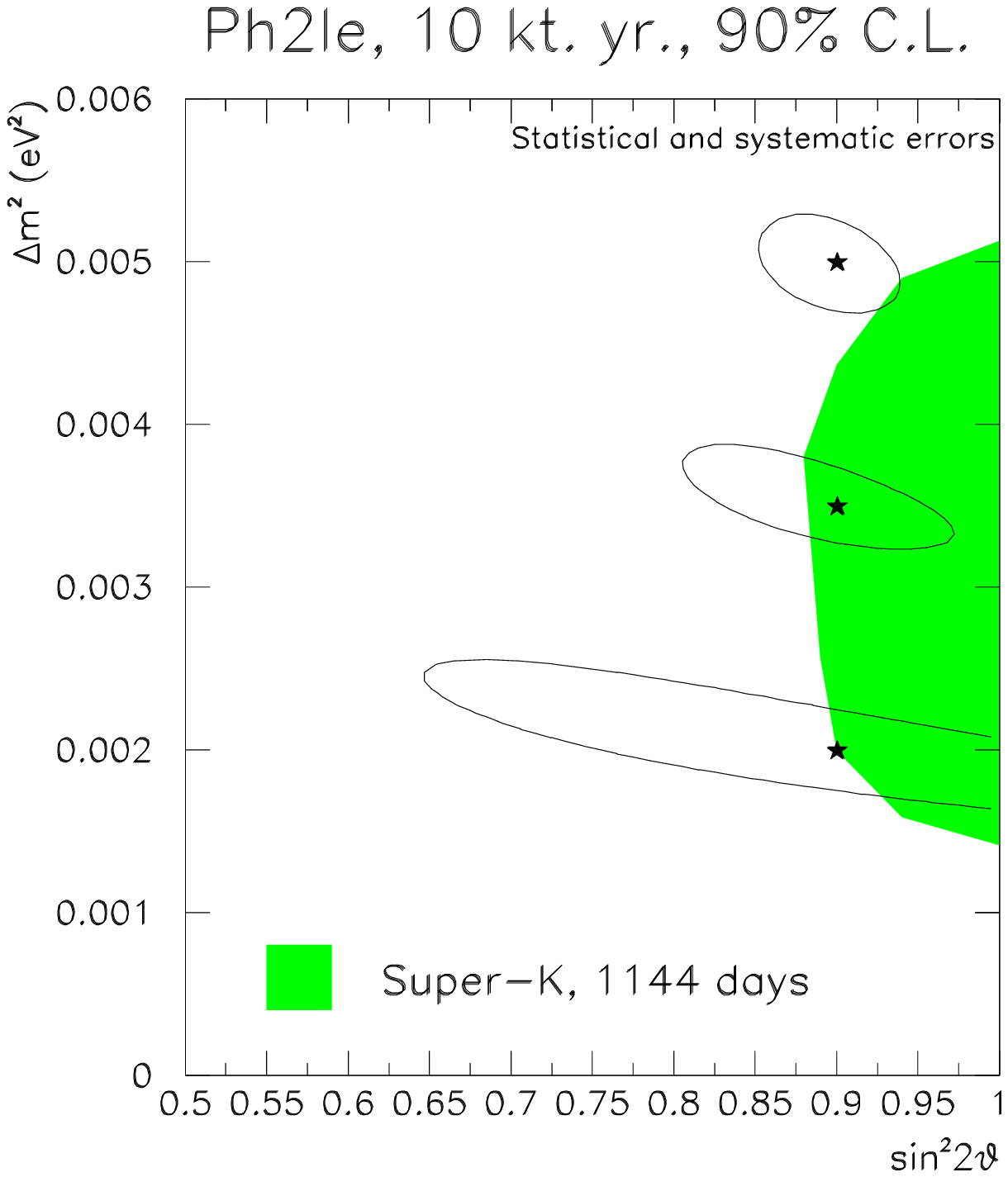}}
\caption{\label{fig:ccenergy} 
{\bf Left top}: \numu\ charged current total energy
distribution oscillated and unoscillated together for the three
different neutrino beam spectra for \dmsq$=3\cdot10^{-3}$ and
\sintwo$=0.8$. Two years of running at nominal intensity have been
assumed. {\bf Left bottom}: 
Ratio of the oscillated and unoscillated energy spectra. A dip
can clearly be seen when using the low energy beam. This is the
strongest indication for neutrino oscillations.
\label{fig:ccerr} {\bf Right}:
The plot shows the 90\% C.L. allowed regions from fits to reconstructed
CC energy distributions. These distributions are generated with specific
oscillation parameters \dmsq, \sintwo\ (indicated by stars on the plot). 
The Super-K 90\% C.L. allowed region is also shown.}
\end{figure}

The MINOS beam has been designed to use a system of two parabolic horns
whose position relative to the target can be adjusted to select a band
of neutrino energies.  Figure~\ref{fig:ccenergy} shows the number of
\numu\ CC events per kiloton of detector per year as a function of
neutrino energy for three configurations of the beam: high, medium and low
energy.  The beam energy can be chosen to give maximum sensitivity to
the region of \dmsq\ to be investigated.  It can be seen that beams
peaking below 2-3 GeV give essentially negligible event rates, thus
setting a lower limit of around 8x10$^{-4}$eV$^2$ on any detectable
oscillations.

The disappearance of \numu\ from the beam as a function of energy will
be reflected in dips in the measured \numu\ CC energy distribution.
\numu\ CC events can be selected with high purity and small losses by a
cut on event length, exploiting the fact that a penetrating muon is
present in the final state. Figure
\ref{fig:ccenergy} shows the measured CC total energy distribution at
Soudan compared with that expected from extrapolation of the near
detector energy distribution for the three different beam configurations
from a two year run\cite{minos-blessed}.  
The plot includes all resolution and selection
effects.  The dips due to the oscillation are most visible with the low
energy beam.  \dmsq\ is given by the position of the minimum of the dip
and \sintwo\ by the magnitude of the dip.  Figure \ref{fig:ccerr} shows
the confidence level contours in \dmsq-\sintwo\ from fits to 
distributions generated with \sintwo=0.8 and different values of \dmsq.  

Oscillations of \numutonutau\ or \numutonue\ result in the rate of
\numu\ CC events being reduced relative to that expected from the near
detector rate and being replaced by \nutau\ or \nue\ events.  
\numu\ CC events are characterised as ``long'' events with a
penetrating muon whereas \nutau\ and \nue\ CC events, along with neutral
current (NC) events, produce only hadronic or electromagnetic showers
and are thus characterised as ``short'' (apart from the 17\% of $\tau$
decays to $\mu\nu\nu$).  The result of oscillations is to increase
the ratio of short to long events (NC/CC ratio).  
The measurement of this ratio in the near and far detector is therefore a
very sensitive indication of oscillations and gives the measurement of
oscillation probability with the smallest systematic errors.  The
\nue- appearance measurement is the third important measurements
done with MINOS. 

Combining all these measurements allows the
explicitly calculate of the oscillation probabilities for \numutonutau,
\numutonue\ and \numutonust.

\subsection{The OPERA Experiment}

OPERA\cite{opera_proposal} is designed as an appearance search for \numutonutau\
oscillations. It is located in the Gran Sasso Laboratory in the CNGS
neutrino beam from the CERN SPS. The detector is based on a massive
lead/emulsion target. The nuclear emulsions are exploited for the direct
observation of the decay of the $\tau$-lepton, produced in \nutau\
charged current interaction.

While the NuMI beam at Fermilab is optimised to see the dip in the 
\numu\ charged current energy distribution, the CNGS beam has been
optimised to detect the appearance of \nutau\ events. Its average energy
is much higher in order to be above the energy threshold for $\tau$ production.
The neutrino beam is generated from a 400 GeV proton beam 
on a graphite target($4.5\times10^{19}$ pot/year) \cite{cngs_tdr}. 
The pions and kaons produced in this beam are focused with a
magnetic horn and reflector into the 900m decay pipe.
The expected neutrino spectrum at Grand Sasso is optimised for the
$\nu_\tau$ appearance measurement and peaks at around 20 GeV.
It is predominately a \numu\ beam with negligible
contaminations from electron or $\tau$ neutrinos.

OPERA aims to detect
the \nutau\ charged current events by identifying the $\tau$ produced
through its decay kink. OPERA is an evolution of the Emulsion Cloud 
Chamber (ECC) technique, which combines high precision tracking in the
emulsion with a passive material to provide the large target mass. 

An ECC cell consists of a $1$~mm thick lead plate
followed by a thin film made of two $50~\mu$m thick emulsion layers separated 
by a $200~\mu$m plastic base.  
The basic detector unit is obtained by stacking 
$56$ cells to form a compact brick ($10.2\times 12.7
\mathrm{cm}^2$ transverse section, $10~X_0$ length, $8.3$~kg weight), 
which will be assembled into walls. 

After each wall, an electronic tracking detector is used to select the
brick where the neutrino interaction occurs. This tracking detector 
follows very closely that of MINOS, with
scintillator strips coupled to wave-length shifting fibres and
multi-anode PMT readout, where the selected PMTs are the same as those used in
MINOS near detector. A wall of bricks and two planes of electronic
detectors constitute one OPERA module, whose structure is then
repeated $24$ times to form a supermodule, also including a
downstream muon spectrometer. Each muon spectrometer consists of a
dipolar magnet ($1.55$~T in the tracking region), made of two vertical 
walls of iron layers, interleaved with RPC detectors and drift
tubes, placed in front and behind the magnet as well as between the
two walls. Three supermodules form the full OPERA detector, for a total
target mass of about $1.8$~ktons.

The data from the electronic detectors, which will be analysed
quasi-online, are used to select bricks where neutrino
interactions occur and which will be removed by an automated system. 
Additional bricks might need to be
removed for a complete reconstruction of the interesting events. Events
selected as $\tau$ candidates will then be sent to dedicated scanning
stations, where further studies can be performed to achieve the
required background suppression. The expected rate of $\nu_\mu$CC
events is of about $30$ events per day.

The $\tau$ decay channels considered by OPERA are listed in
Table \ref{tab:opera_decay_channels}, 
together with the corresponding
$\nu_\tau$ detection efficiencies and the expected number of background
 events, which are mostly from charm production, large angle muon scattering
   and hadron re-interactions. New studies are
underway to improve the experiment sensitivity by adding also the
$\tau\rightarrow \rho\nu$ channel ($23.5\%$ B.R.) 
to the decay modes considered.
The expected number of $\tau$ events per year is given in
Table \ref{tab:opera_expected_ntau} as a function of possible values of \dmsq.
\cite{opera_nashville}

\begin{table}
\centerline{
\begin{tabular}{|c||c|c|}
\hline
Decay mode & $\epsilon(\nu_\tau CC)$(\%) & $N_{BKGD}/$~year\\
\hline\hline
$\tau\rightarrow e$   & $3.7$ & $0.04$\\
$\tau\rightarrow\mu$  & $2.7$ & $0.03$\\
$\tau\rightarrow h$   & $2.3$ & $0.05$\\
\hline
\end{tabular}
\hspace{2cm}
\begin{tabular}{|c|c|}
\hline
\dmsq (eV$^2$) & $N_\tau$/year\\
\hline\hline
$1.5\times 10^{-3}$ & $0.82$\\
$2.5\times 10^{-3}$ & $2.82$\\
$3.2\times 10^{-3}$ & $3.66$\\
\hline
\end{tabular}}
\caption{\label{tab:opera_decay_channels} 
{\bf Left}: OPERA efficiency for the $\tau$ decay
channels considered and the corresponding expected number of background
events per year.  The efficiencies also include the branching
ratio for each decay channel. \label{tab:opera_expected_ntau} 
{\bf Right}:
Expected number of $\tau$ events observed in OPERA per year as a
function of possible values of \dmsq.}
\end{table}

\begin{figure}
\centerline{
\includegraphics[width=0.45\textwidth]{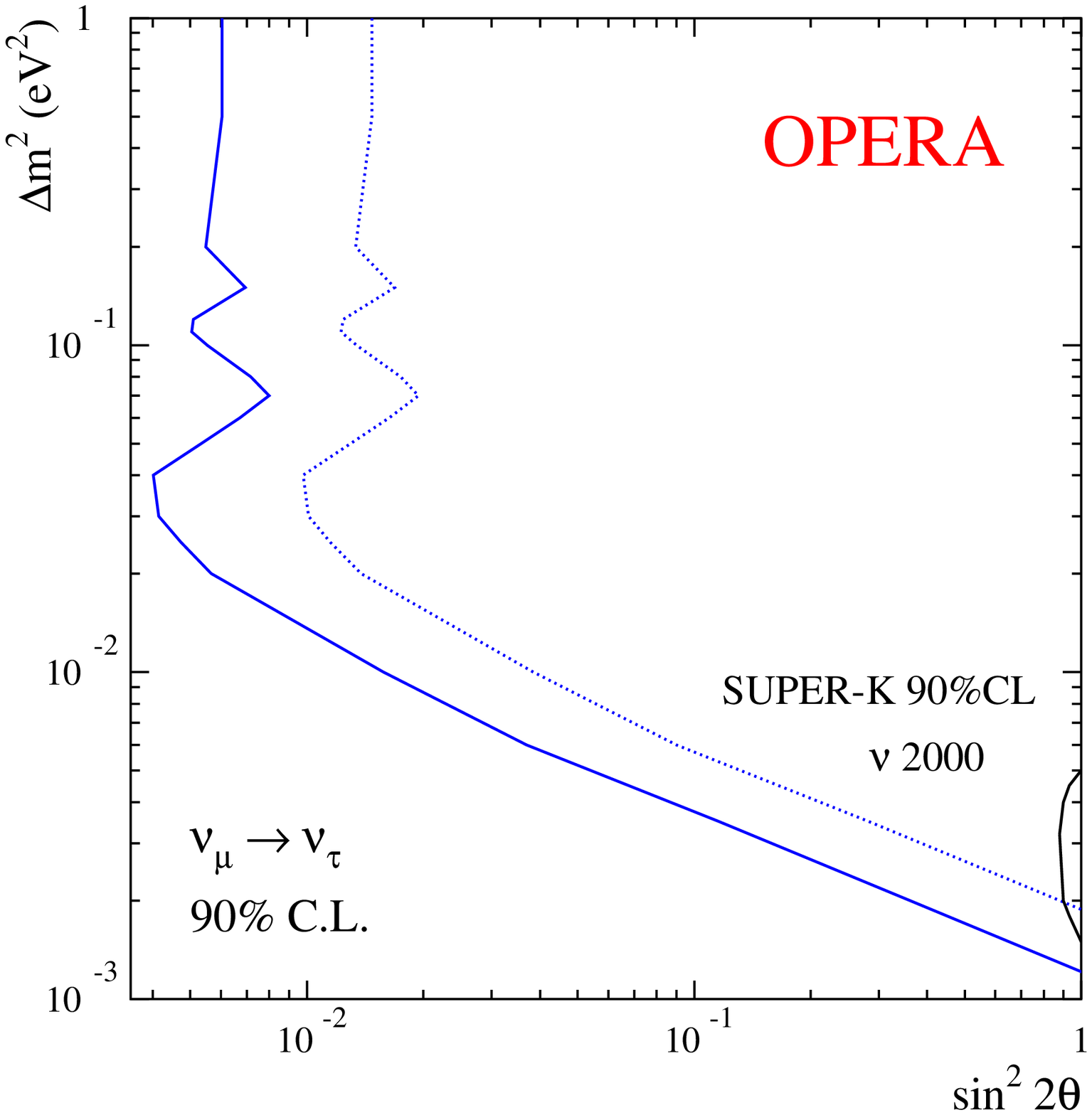}\hspace{1cm}
\includegraphics[width=0.45\textwidth]{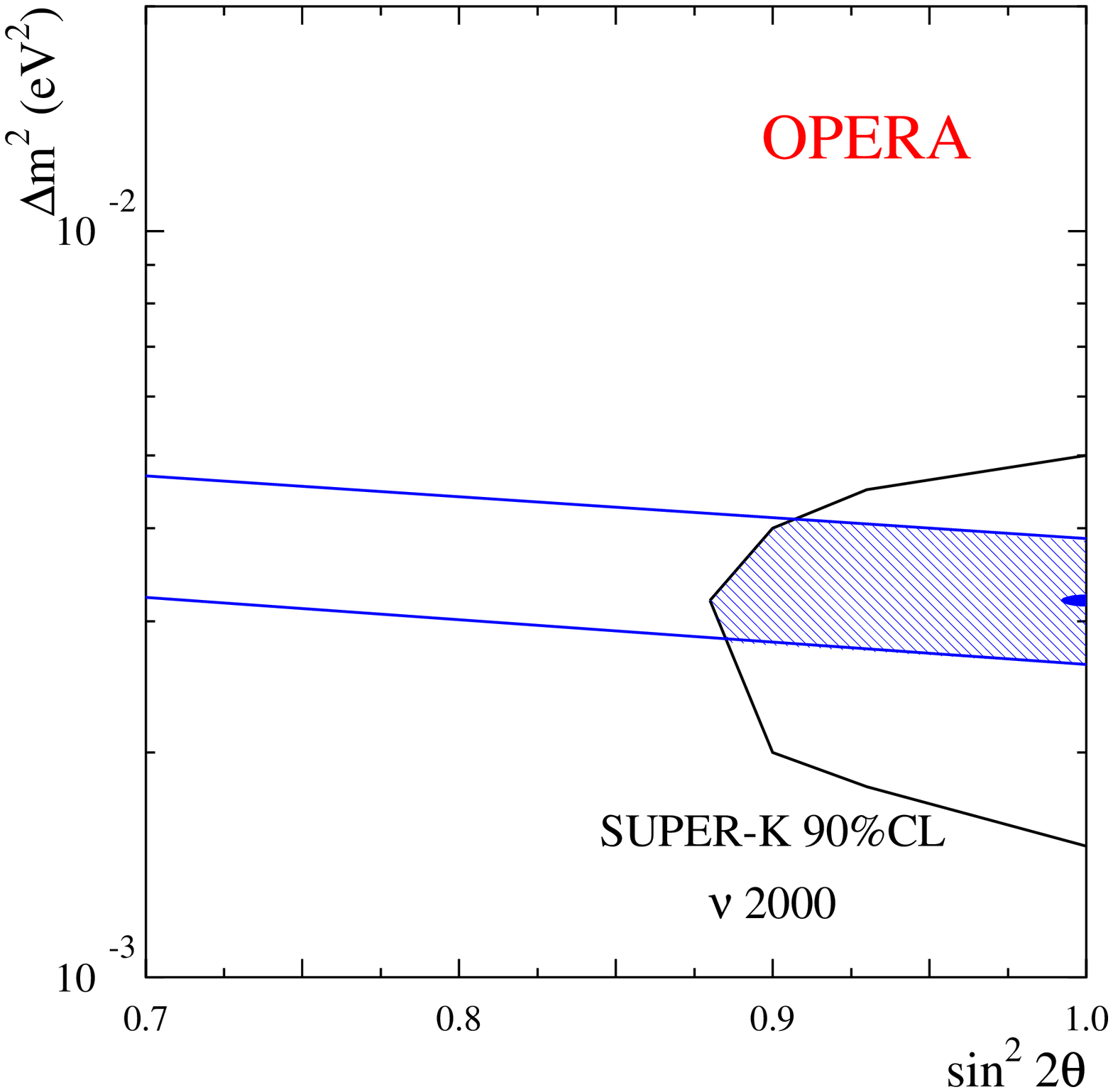}}
\caption{\label{fig:opera_sens} 
{\bf Left}: 
Expected upper limit at the 90\% C.L. from OPERA with no oscillation
signal after 2 and 5 years of running.
\label{fig:opera_allowed} {\bf Right}:
Possible 90\%C.L. allowed region for the oscillation parameters from OPERA
after 5 years of running assuming an oscillation signal.
}
\end{figure}

If no signal is observed, the average upper limit at $90\%$~C.L. which would be
obtained by OPERA is shown in Figure \ref{fig:opera_sens} for
$2$ and $5$ years of exposure respectively. On the other hand, if
$\nu_\tau$ events are actually observed, a measurement of \dmsq can 
be performed. Assuming maximal mixing and \dmsq$ = 3.2\times
10^{-3}~\mathrm{eV}^2$, the $90\%$~C.L. allowed region for the oscillation
parameters as determined by OPERA after $5$ years of data taking is shown in
Figure \ref{fig:opera_allowed}.

OPERA also plan to perform a $\nu_\mu\rightarrow\nu_e$ oscillation
analysis, by looking for $\nu_e$ appearance above the $\sim 1\%$ intrinsic
$\nu_e$ contamination, and to search for neutrino oscillations by
studying the ratio between NC and CC
interactions. However the
study of the systematic uncertainties for those analyses is likely to
be somewhat limited by the lack of a near detector.

\subsection{KamLAND}

The KamLAND experiment\cite{kamland} has just started to take data.
This experiment will not detect muon neutrinos from an accelerator, but
measure the flux and energy spectrum of 1-8 MeV $\bar{\nu}_e$ emitted by
the nuclear reactors at several Japanese power stations, at a distance
between 150 and 210 km from the detector. This experiment will therefore
not make additional contribution to the atmospheric neutrino problem,
but investigate the solar neutrino oscillation scenario.  The estimated
sensitivity of KamLAND is such that at 90\%CL and after three years of
data taking, the experiment should be able to cover the entire domain
defined by the LMA solution (down to \dmsq$\sim
4\times10^{-6}\mathrm{eV}^2$). If oscillations are observed, the
parameters should be determined to a precision of 20\% at the 99\%
CL. This is relevant not only for the solar neutrino problem, but also,
if LMA is confirmed, may allow to measure CP violation in the lepton
sector.
 
\section{The Future}

This initial round of long baseline neutrino experiments will
establish once and for all the existence of neutrino oscillations and
measure the parameters $\Delta m^2_{23}$ and $\sin^22\theta_{23} $ 
with a 5\% to 20\% precision.
However, further progress in the field can only be made using even more
intense neutrino beams with energies and oscillation length tuned to the
question to be answered. It is especially interesting to investigate the
possibility of CP violating effect in the lepton sector of the Standard
Model. The way forward here might be to either build a conventional
neutrino beam with an even higher flux (``Super-Beam'') or to develop a
neutrino source based on a muon storage ring (``Neutrino Factory'').

\subsection{Super-Beams}

There are currently no constrain on the parameter $\sin^22\theta_{13}$,
which describes the sub-dominant mixing of muon neutrinos to electron 
neutrinos in the atmospheric neutrino oscillation scenario. The sensitivity of
current experiments to this parameters is quite limited. An improvement
in sensitivity can be achieved by so-called ``Super-beams''.

A Super-beam is a conventional neutrino beam with a largely increased
neutrino flux. Ideas for such machines are currently being discussed at
several accelerator centres\cite{superbeam}. The most concrete proposal
is currently considered in Japan\cite{jhf}, in which
a high intensity narrow band neutrino beam is produced by secondary
pions created by the high intensity proton synchrotron at JHF. The
neutrino energy could be tuned to the oscillation maximum at 1 GeV for a
baseline length of 295 km towards the Super-Kamiokande detector. Its
energy resolution and particle identification enable the reconstruction
of the initial neutrino energy. The physics goal of the first phase of
the proposal is an order of magnitude better precision in the
\numutonutau\ oscillation measurement and a factor of 20 more
sensitivity in the \numutonue\ appearance mode than conventional
currently approved experiments. In the second phase, an
upgrade of the accelerator from 0.75 MW to 4 MW in beam power and the
construction of a 1 Mton Hyper-Kamiokande detector are envisaged. This
would lead to an order of magnitude improvement in the \numutonue\
oscillation sensitivity and may make the measurement of a CP violating
phase down to $10^\circ-20^\circ$ possible.

\subsection{Neutrino Factories}

\begin{figure}
\centerline{
\includegraphics[width=0.5\textwidth]{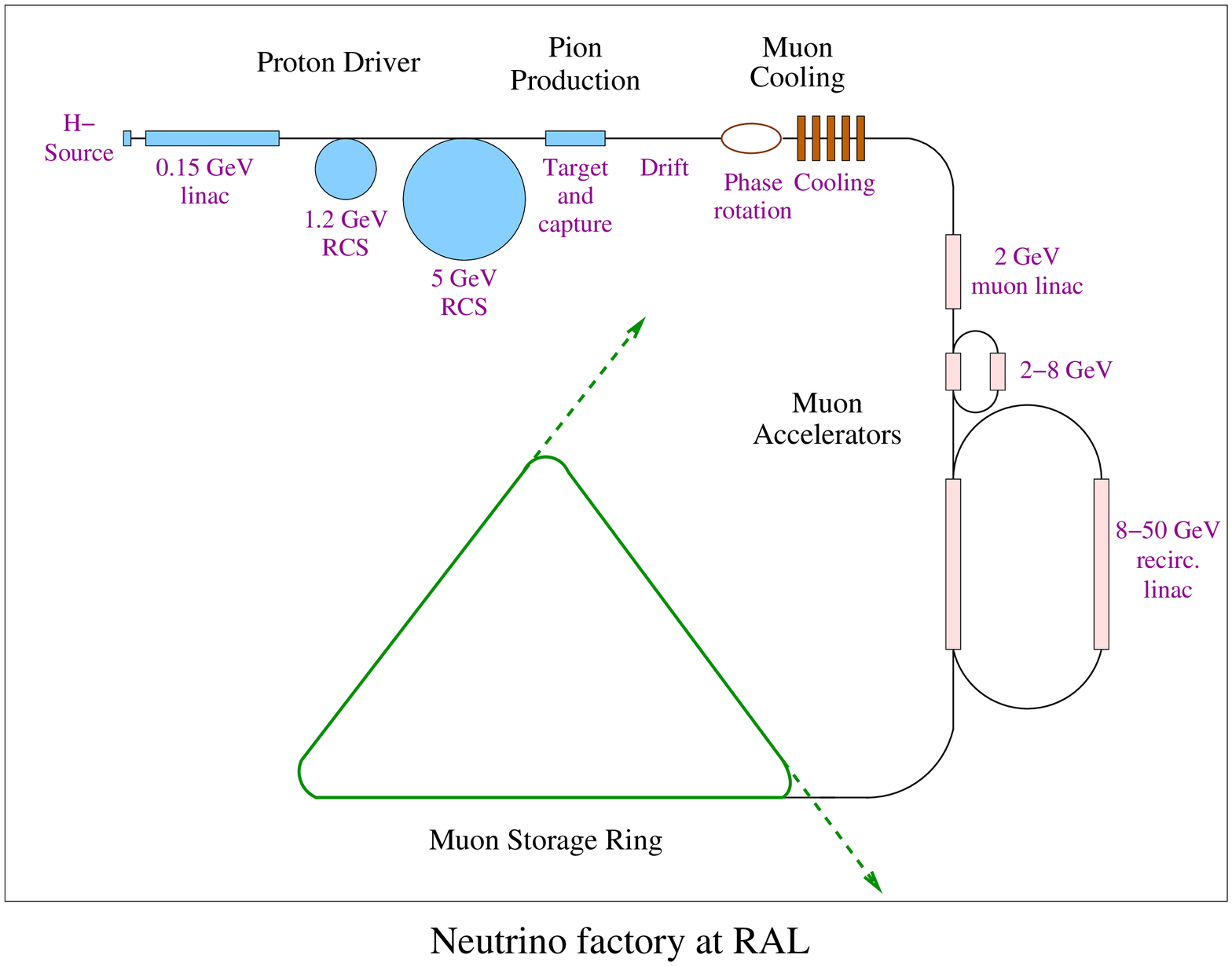}
\hspace{0.05\textwidth}
\includegraphics[width=0.45\textwidth]{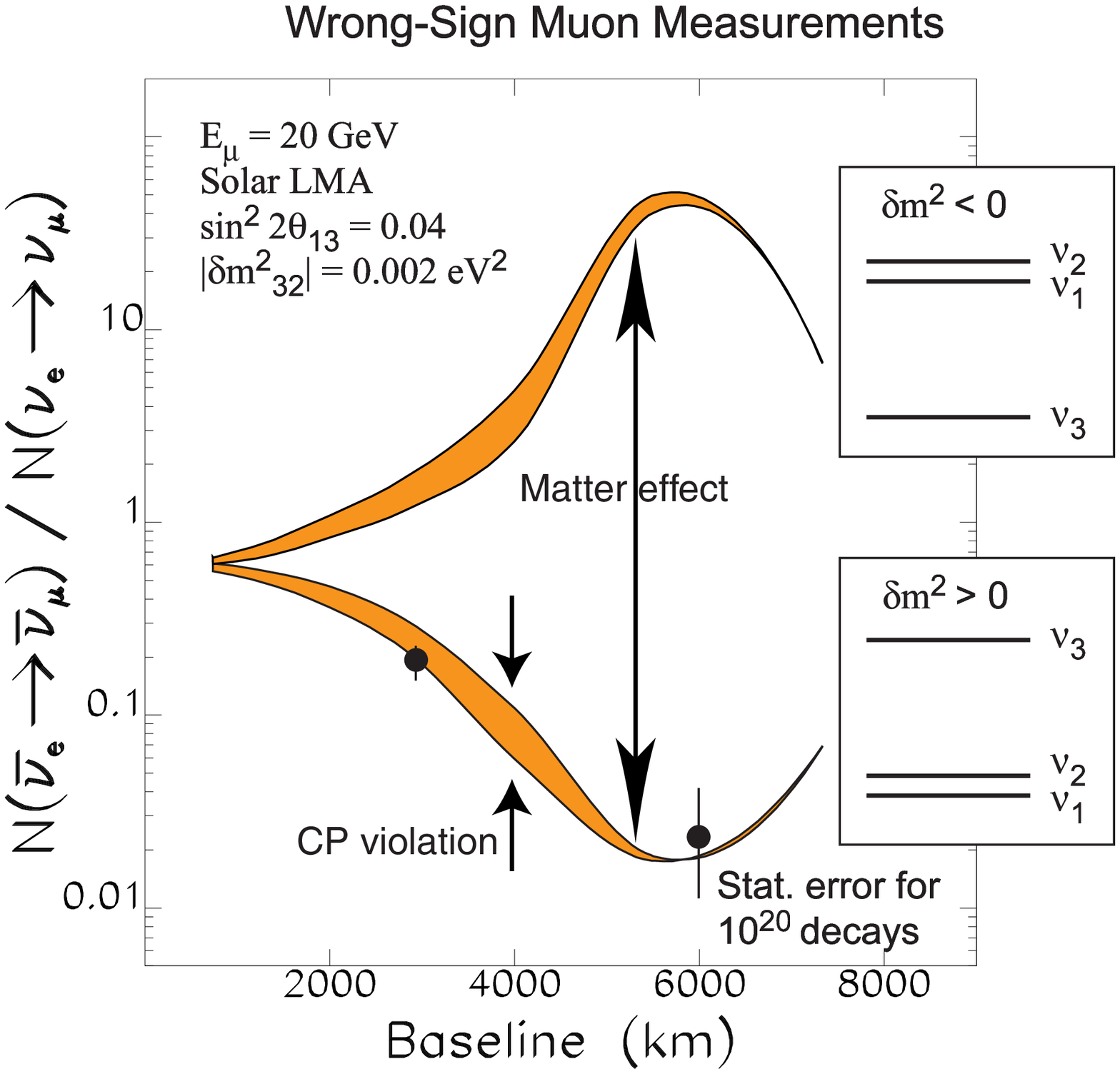}
}
\caption{\label{fig:nufact}{\bf Left}: 
Possible layout of a neutrino factory. An
intense beam of protons is dumped onto a target. The resulting pions
decay and produce muons. This beam of muons is cooled by passing it
repeatedly through a liquid Hydrogen absorber and re-accelerating
it. The muon beam is than accelerated by a recirculating linear
accelerator to its final energy of 20-50 GeV. This muons are put into a
storage ``ring'' with two or three straight section. The neutrino beam
comes from the muons decaying in the straight sections of the
ring. A ``bow-tie'' or triangular layout would allow service to
experiments in different locations and baselines.
{\bf Right}:
This diagram shows the CP violating effect
on the measured rate of electron neutrinos oscillating to muon neutrinos
compared to the anti-particle oscillations for certain mixing
parameters. An exact measurement of this ratio allows not only a
measurement of the CP violating phase at moderate baselines, but also 
determines the mass ordering of the neutrino states.}
\end{figure}

The ultimate weapon to measure mixing parameters and the CP-odd
phase in the neutrino mixing matrix is a neutrino factory. This is
a revolutionary idea which may possibly lead to the most intense
neutrino sources. In a conventional neutrino beam protons are dumped
onto a target to produce pions. These pions decay to produce muon
neutrinos and muons. Contrary to conventional beam the where the muons
are absorbed and the neutrino beam is made from the pion decay
neutrinos, the muons are captured in a neutrino factory.
This muon are then accelerated and put into a storage ring to decay
(see Fig.\ \ref{fig:nufact}).

There have been a number of feasibility studies in the US, Japan and
also at CERN, which try to address the main technological problems of
such a machine.\cite{nufact_study} Some of the technological problems
that must be solved in order to achieve the full physics potential are
listed below.
\begin{itemize} 
\item build a proton target that can withstand a 4 MW beam, 
\item reduce the emittance of the resulting muon beam so that it fits into
an accelerator, and
\item reduce the costs so that one can actually afford it. 
\end{itemize}

The muon decay is a pure weak process and well understood. This leads to
a much simpler neutrino beam containing significant and precisely known
amount of electron and muon neutrinos. This eliminates the errors
coming from the hadronic pion production uncertainties in conventional
neutrino beams.  A figure of merit for a neutrino factory is shown
Fig.\ref{fig:nufact}. A neutrino factory will be able to determine the
ordering of the neutrino mass eigenstates due to matter effects and also
lead to a measurement of the CP violating phase, if the other neutrino
parameters are right (solar LMA solution and non-vanishing
$\sin^22\theta_{13}$).\cite{nufact}

\section{Summary}
Much progress has been made since the first suggestions that the
atmospheric neutrino anomaly could be explained by neutrino
oscillations. K2K is the first of a new class of long baseline
experiments which try to confirm the oscillation hypotheses. However,
it's sensitivity is limited and only the new long baseline experiments
like MINOS and OPERA will deliver the final proof that the atmospheric
neutrino deficit is due to neutrino oscillations. Current SNO data
suggest that the LMA solution explains the solar neutrino
problem\cite{sno}. KamLAND, which recently started data taking should
therefore be able to measure and confirm this scenario.  More
interesting results, like a measurement of the so far unknown parameter
$\sin^22\theta_{13}$, and a possible determination of CP violating
parameters in the neutrino sectors will lay in the hands of future
machines and require significant improvements in accelerator technology
and detector size.

\section*{Acknowledgements} 

I would like to thank the KamLAND, K2K, MINOS, and OPERA collaborations
for providing material for this talk. Special thanks to N.~Tagg and
J.~Cobb for carefully reading this manuscript and the talk.  This talk
was sponsored by the Conference \& Schools Committee of the Particle
Physics sub-department, University of Oxford.

\end{document}